\documentclass{mn2e}
\usepackage{natbib}
\usepackage{graphicx}
\usepackage{pslatex}

\newcommand{\ltappeq}{\raisebox{-0.6ex}{$\,\stackrel
{\raisebox{-.2ex}{$\textstyle <$}}{\sim}\,$}}
\newcommand{\gtappeq}{\raisebox{-0.6ex}{$\,\stackrel
{\raisebox{-.2ex}{$\textstyle >$}}{\sim}\,$}}

\begin{document}

\title[Flat spectra of self-absorbed jets]{The flat synchrotron spectra of partially self-absorbed jets revisited}

\author[C.R. Kaiser]{C. R. Kaiser\thanks{crk@soton.ac.uk}\\ 
School of Physics \& Astronomy, University of Southampton, Southampton SO17 1BJ
}

\maketitle

\begin{abstract}
Flat radio spectra with large brightness temperatures at the core of AGN and X-ray binaries are usually interpreted as the partially self-absorbed bases of jet flows emitting synchrotron radiation. Here we extend previous models of jets propagating at large angles to our line of sight to self-consistently include the effects of energy losses of the relativistic electrons due to the synchrotron process itself and the adiabatic expansion of the jet flow. We also take into account energy gains through self-absorption. Two model classes are presented. The ballistic jet flows, with the jet material travelling along straight trajectories, and adiabatic jets. Despite the energy losses, both scenarios can result in flat emission spectra, however, the adiabatic jets require a specific geometry. No re-acceleration process along the jet is needed for the electrons. We apply the models to observational data of the X-ray binary Cygnus X-1. Both models can be made consistent with the observations. The resulting ballistic jet is extremely narrow with a jet opening angle of only 5". Its energy transport rate is small compared to the time-averaged jet power and therefore suggests the presence of non-radiating protons in the jet flow. The adiabatic jets require a strong departure from energy equipartition between the magnetic field and the relativistic electrons. These models also imply a jet power two orders of magnitude higher than the Eddington limiting luminosity of a 10\,M$_{\odot}$ black hole. The models put strong constraints on the physical conditions in the jet flows on scales well below achievable resolution limits.
\end{abstract}


\section{Introduction}

The centres or cores of many AGN show a flat radio spectrum in the sense that for the flux density as a function of frequency $\nu$ we observe $F_{\nu} \propto \nu^{\alpha}$ with $\alpha \sim 0$. The high surface brightness temperature associated with these spectra suggests a synchrotron origin of the emission. Observations with high spatial resolution reveals that the flat spectrum arises in the base of jet flows which continue to much larger scales \citep[for a review see][]{tc91}. Similar flat or inverted ($\alpha > 0$) radio spectra are also observed in X-ray binaries in the low-hard state \citep[e.g.][]{rf01}. If optically thin, the flat synchrotron spectrum would imply a power-law energy distribution of the radiating relativistic electrons with a slope of unity. Such a distribution is very unlikely to arise for the usually assumed mechanism for the acceleration of the electrons at shock fronts \citep[e.g.][]{ab78}.

A magnetised plasma containing very energetic electrons with a power-law energy distribution will produce a power-law spectrum at high frequencies. The slope of the spectrum, typically $\alpha < 0$, is determined by the slope of the energy distribution. However, below a critical frequency the radiating electrons will re-absorb some of the photons. In this self-absorbed, optically thick regime the spectrum has a power-law slope of $5/2$, independent of the slope of the electron energy distribution \citep[e.g.][]{rl79}. The spectrum of a uniform, self-absorbed synchrotron source therefore shows a pronounced peak. \citet{bk79} pointed out that in a jet the plasma conditions are changing along the flow and therefore the peaks of the self-absorbed spectra of different parts of the jet can occur at different frequencies. If the plasma conditions change such that the spectra peak at the same level, then the overall spectrum, observed with a spatial resolution insufficient to resolve the individual parts of the jet, will be flat. Their model has become the standard tool for interpreting observations of flat radio spectra from jetted sources. 

In the \citet{bk79} model the jet is assumed to have a conical geometry, i.e. the velocity with which the jet is expanding sideways, is constant. The bulk velocity of the jet material along the jet axis is also assumed to be constant. The magnetic field is assumed to be directed perpendicular to the jet axis and `frozen' into the jet plasma. Adiabatic losses of the electrons are mentioned by the authors, but are assumed to be replenished by an unknown, continuous re-acceleration process along the entire jet. The same assumption is made for radiative energy losses associated with the emission of synchrotron radiation. The subsequent model of \citet{am80} includes a simplified treatment of energy losses of the electrons due to adiabatic expansion and radiative processes. It also allows for more confined jets, i.e. the jets are not necessarily conical. With these assumptions, the model is unable to produce a flat emission spectrum. A similar model was developed by \citet{hj88}. They consider adiabatic, but not radiative, energy losses of the electrons. The jet geometry is again conical, but they also investigate a more confined jet. The model can predict flat spectra, but \citet{hj88} point out that these may only arise under special circumstances, particularly in the case of confined jets. The model of \citet{gm98} includes a detailed treatment of the energy losses of the relativistic electrons, but it concentrates only on the optically thin part of the spectrum of jets propagating close to the line of sight for which numerical solutions are presented. The perhaps most comprehensive study of jet emission models is that of \citet{sr82} which includes the effects of energy losses on the electron population, but neglects the effects of self-absorption on the electron energy spectrum.

In this paper we extend the previous models by including adiabatic and radiative energy losses and gains (due to absorption) for the electrons as well as investigating various possibilities for the evolution of the magnetic field. We consider two distinct cases: The ballistic and the adiabatic jet models. In the ballistic case the jet material follows straight trajectories and does not behave like a fluid, because individual fluid elements do not interact with each other. In many ways this model is similar to the \citet{bk79} model, but we show that because of self-absorption effects we do not need to invoke a re-acceleration process to achieve flat emission spectra. in the adiabatic jet model the relativistic electrons suffer from adiabatic energy losses as well as radiative losses. Again we show that the models can produce flat spectra without re-acceleration of the electrons, but only for a very specific jet geometry. The emphasis of our treatment is on the construction of analytical models and so we concentrate on jets propagating at large angles to the line of sight, i.e. the viewing angle is larger than the inverse of the Lorentz factor of the jet flow.

In Section \ref{conical} we briefly discuss the basic properties of our jets in terms of their geometry, the evolution of the magnetic field and that of the relativistic electrons. We present the first fully analytical solution of the equations governing partially self-absorbed synchrotron emission from a jet in Section \ref{radiation}. Section \ref{simple} summarises the model results for the case without radiative energy losses as studied in many previous models. In Section \ref{cutoff} we develop the formalism for including radiative energy losses in the model and the resulting spectra are discussed in Section \ref{losses}. We apply the model to the data obtained for Cygnus X-1 in Section \ref{obs} and derive the properties of this jet. Finally, we summarise our conclusions in Section \ref{conc}.

\section{The model}

In this Section we derive the emission properties of partially self-absorbed jets neglecting radiative energy losses of the relativistic electrons. Note that we are concentrating on jets at comparatively large angles to our line of sight, $\vartheta$. As we will point out further down this greatly simplifies the determination of the optical depth of the jet material. 

\subsection{The basic jet properties}
\label{conical}

\subsubsection{Jet geometry and velocity}

We take the $x$-axis as the centre of a jet that is rotationally symmetric about this axis. The geometrical shape of the jet is then given by a one-dimensional function $r(x)$ defining the jet radius with respect to the $x$-axis. Analogous to previous work we parameterize this function as $r(x)= r_0 \left(x / x_0 \right)^{a_1} = r_0 l^{a_1}$, where $x_0$ is an arbitrary position along the $x$-axis defining the dimensionless coordinate $l$ and $r_0$ is a constant scaling factor. The value of the exponent $a_1$ depends on the details of the confinement of the jet. Confinement by external pressure is the simplest mechanism \citep{br74}, but can lead to problems with the collimation of the jet \citep{bbr84}. Confinement by magnetic fields has also been suggested by various authors, but it is unlikely that magnetic fields alone, without additional gas pressure, can collimate the jet on large scales \citep{mb95}. For our purposes here we do not need to specify the details of the jet confinement and we will assume that $0\le a_1 \le1$. In principle one could also envisage highly overpressured jets with an accelerating expansion rate, i.e. $a_1 > 1$. However, the pressure in such jets would fall very rapidly and they would quickly evolve to a situation where $a_1 \le 1$.

The extreme case of a highly overpressured jet is that of a jet flow expanding into a (near) vacuum. In such a ballistic jet, as opposed to the adiabatic, confined jet discussed above, the jet freely expands in the direction perpendicular to the jet axis. In this process random, `thermal' energy of the jet material is converted to ordered, kinetic energy associated with the sideways expansion. The random energy of the electrons giving rise to the synchrotron emission is reduced by this adiabatic expansion.
However, in Section \ref{cyg} we apply the ballistic jet model to the observations of the jet in Cygnus X-1. There we will find a very small opening angle for the ballistic jet implying very small adiabatic expansion losses. Therefore we can assume that in the limiting case of a ballistic jet studied here the relativistic electrons do not suffer energy losses other than those associated with radiation processes. 

We assume in this paper that the velocity of the jet material along the jet axis, $v_{\rm j}$, is constant. While this is justified in the case of the ballistic jet, the confined, adiabatic jets can be accelerated, for example, by a pressure gradient in the confining medium. In the model of \cite{br74} the Lorentz factor of the bulk velocity is proportional to $p_{\rm x}^{-1/4}$, where $p_{\rm x}$ is the pressure of the external medium. As long as the external pressure gradient is shallow, the Lorentz factor of the jet flow will be only a very weak function of the position along the jet axis. Similar arguments hold for a magnetically confined jet. The constant bulk velocity of the jet also implies that a given volume element $\Delta V$ travelling with the jet flow will only expand sideways according to $\Delta V \propto r^2$. A constant jet velocity also simplifies the model greatly as we can ignore the effects of varying length contraction along the jet axis \citep{gk04}.

\subsubsection{Magnetic field}

The strength of the magnetic field changes during the sideways expansion of the jet material. In general, we parameterize the evolution of the magnetic field as $B = B_0 l^{-a_2}$. For flux freezing of the magnetic field and using flux conservation we have that the field component parallel to the jet axis, $B_{\parallel}$, is proportional to $r^{-2}$. Also, the magnetic field components perpendicular to the jet axis, $B_{\perp}$, are proportional to $r^{-1}$. For an initially mixed field, $B_{\perp}$ will always become the dominant component and so $a_2 =  a_1$. The perpendicular magnetic field may also contribute to the confinement of the jet. For completeness we also consider a purely parallel configuration of the magnetic field with $a_2 = 2a_1$. Finally, if the magnetic field is constantly tangled by turbulent motions in the jet material on scales smaller than the jet radius, then it can remain isotropic and it behaves like a relativistic fluid with $B = B_0 l^{-4 a_1 / 3}$ and $a_2 = 4 a_1 /3$ \citep[e.g.][]{hb00}. This is analogous to the behaviour of the magnetic field in an isotropic expansion \citep[e.g.][]{ml94}, but is clearly incompatible with flux freezing. The last case of a permanently isotropic field cannot be realised in the ballistic jet as it would require that the jet material behaves like a fluid.
 
 \subsubsection{Relativistic electrons}
 
In order to produce synchrotron emission the jets must contain a population of relativistic electrons. We assume that the latter has a power-law energy distribution of the form
\begin{equation}
N(E)\,{\rm d}E = \kappa E^{-p} \, {\rm d}E,
\end{equation}
where $E$ is the electron energy, $E=\gamma m_{\rm e} c^2$, and $\kappa$ is a scaling independent of $E$. $\gamma$ is the Lorentz factor associated with the relativistic motion of the electrons. In this section we do not impose a high-energy cut-off on the energy distribution and we neglect radiative energy losses. Even so the energy distribution of the electrons changes as the jet expands. We represent the evolution of the electron distribution by setting $\kappa = \kappa _0 l^{-a_3}$. 

For a given volume of jet material $\Delta V$ particle conservation demands that
\begin{equation}
\Delta V \kappa \gamma ^{-p} \, {\rm d}\gamma = \Delta V_0 \kappa _0 \gamma _0^{-p} \, {\rm d}\gamma_0,
\label{dist}
\end{equation}
where all quantities with subscript `0' refer to their values at $x=x_0$. Therefore for the ballistic jet we have $a_3 = 2 a_1 = 2$. 

For the adiabatic jet we need to include energy losses due to the jet expansion. Since most of the electrons are highly relativistic we have \citep[e.g.][]{ml94}
\begin{equation}
\frac{\partial \gamma}{\partial t} = - \frac{1}{3} \gamma \frac{\partial \ln \left( \Delta V \right)}{\partial t},
\label{adiabat}
\end{equation}
which has the solution 
\begin{equation}
\gamma = \gamma _0 \left( \frac{\Delta V}{\Delta V_0} \right)^{-1/3}, 
\label{adsol}
\end{equation}
and it follows that
\begin{equation}
\frac{\partial \gamma _0}{\partial \gamma} = \left( \frac{\Delta V}{\Delta V_0} \right)^{1/3}.
\end{equation}
Re-arranging equation (\ref{dist}) and substituting yields
\begin{equation}
\kappa \gamma ^{-p} \, {\rm d} \gamma = \frac{\Delta V_0}{\Delta V} \kappa _0 \left( \frac{\Delta V}{\Delta V_0} \right)^{-p/3} \gamma^{-p} \left( \frac{\Delta V}{\Delta V_0} \right)^{1/3} \, {\rm d} \gamma.
\end{equation}
Collecting terms and remembering that $\Delta V \propto r(x)^2$ we find $a_3 = (4+2p) a_1/ 3$.

\subsubsection{Individual models}

On the basis of the discussion above we formulate five individual models distinguished by the magnetic field behaviour. The ballistic jet models, B1 and B2, as well as the adiabatic models, A1 and A2, correspond to a perpendicular and parallel field structure, respectively. The adiabatic model A3 represents the case of an isotropic magnetic field in the jet. The relevant coefficients describing the jet geometry and the behaviour of the magnetic field and relativistic particles are summarised in Table \ref{expo}.

\begin{table}
\begin{tabular}{llccccc}
 & & $a_1$ & $a_2$ & $a_3$ & $a_4$ & $ a_5$\\
 \hline
 ballistic & B1 & 1 & 1 & 2 & $-4-p$ & $\frac{-5}{4+p}$\\[1.5ex]
 	      &	B2 & 1 & 2 & 2 & $-6-2p$ & $\frac{-3}{2p+3}$\\[1.5ex]
 adiabatic & A1 & $a$ & $a$ & $\frac{\left(4 + 2p \right) a}{3}$ & $\frac{-a \left(8 + 7p \right)}{3}$ & $\frac{-3 \left( 3a +2 \right)}{a \left(8 +7p \right)}$\\[1.5ex]
 & A2 & $a$ & $2a$ & $\frac{(4+2p)a}{3}$ & $\frac{-2 a\left( 7 +5p \right)}{3}$ & $\frac{-3 \left(2a +1 \right)}{a \left(7+5p \right)}$\\[1.5ex]
  & A3 & $a$ & $\frac{4 a}{3}$ & $\frac{(4+2p)a}{3}$ & $\frac{-2a \left(5+4p \right)}{3}$ & $  \frac{-5a - 3}{a \left(5 +4p \right)}$\\
 \hline
 \end{tabular}
 \caption{Exponents of the model parameters used in this paper. See text for details.\label{expo}
}
 \end{table}

\subsection{Partially self-absorbed synchrotron emission from a jet}
\label{radiation}

From the expressions for $r$, $B$ and $\kappa$ defined in the previous Section, we can now build a model for the emission from the jet. For this purpose we split the jet into small segments of length ${\rm d} x$ along the $x$-axis. We assume that the segments move along the jet axis at a constant velocity $v_{\rm j} = \beta _{\rm j} c$ corresponding to a Lorentz factor $\gamma _{\rm j}$. The jet axis is at an angle $\vartheta$ to the line of sight of the observer and so the Doppler factor for an approaching (`$-$') or receding (`$+$') jet is $\delta _{\mp} = \left[ \gamma _{\rm j} \left( 1 \mp \beta _{\rm j} \cos \vartheta \right)\right]^{-1}$. The observable monochromatic intensity of one such segment taking into account absorption is
\begin{equation}
I_{\nu} = \delta _{\mp}^3 \frac{J_{\nu}}{4 \pi \chi_{\nu}} \left( 1 - e^{-\chi_{\nu} r(x)} \right),
\end{equation}
where $J_{\nu}$ is the emissivity per unit volume and $\chi_{\nu}$ is the absorption coefficient. For convenience in the development of the model the frequency $\nu$ is measured in the restframe of the jet material. It is related to the observing frequency by $\nu_{\rm ob} = \delta_{\mp} \nu$. Here and in the following we assume that the average path of a photon through the jet has the length $r(x)$. An exact calculation of the radiative transfer of photons through various jet elements would have to take into account relativistic aberration effects. It is therefore complex and impossible in an analytical model. The assumption of an average path length $r$ will not introduce a large error as long as the angle to the observer's line of sight is large. 

The jet segment has a surface area of $2 \pi r(x)\,{\rm d} x$ and so the observable flux density of the segment is given by
\begin{equation}
{\rm d} F_{\nu} = \delta _{\mp}^3 \frac{r(x) J_{\nu}}{2 D^2 \chi_{\nu}} \left( 1 - e^{-\chi_{\nu} r(x)} \right) \, {\rm d} x,
\label{pflux}
\end{equation}
where $D$ is the distance of the jet from the observer. 

Substituting the dimensionless variable $l=x/x_0$, we can express $J_{\nu}$ and $\chi_{\nu}$ in SI units as \citep{ml94}
\begin{eqnarray}
J_{\nu} & = & J_0 \nu^{\left(1-p \right)/2} l^{-a_3-a_2 \left( p+1 \right)/2}\nonumber\\
\chi_{\nu} & = & \chi _0 \nu^{\left(-p-4 \right)/2} l^{-a_3-a_2 \left(p+2 \right)/2},
\label{jk}
\end{eqnarray}
with
\begin{eqnarray}
J_0 & = & 2.3 \times 10^{-25} \left( 1.3 \times 10^{37} \right)^{\left(p-1 \right)/2} c_1(p) \nonumber \\
& & B_0^{\left( p + 1\right)/2} \kappa _0 \,{\rm W\,m^{-3}\,Hz^{-1}}\nonumber\\
\chi_0 & = & 3.4 \times 10^{-9} \left( 3.5 \times 10^{18} \right)^p c_2(p) B_0^{\left( p +2 \right)/2} \kappa _0 \, {\rm m^{-1}},
\label{jk0}
\end{eqnarray}
and  the constants $c_1(p)$ and $c_2(p)$ given by equations 18.49 and 18.74 in \citet{ml94}. Substituting into equation (\ref{pflux}) and integration gives the total flux density of the jet as
\begin{equation}
F_{\nu} = \delta _{\mp}^2 \frac{x_0 r_0 J_0}{2 D^2 \chi_0} \nu^{5/2} \int _{l_{\rm min}}^{l_{\rm max}} l^{a_1+a_2/2} \left[ 1 - e^{-\tau} \right] \, {\rm d}l,
\label{lflux}
\end{equation}
where $l_{\rm min}$ and $l_{\rm max}$ are the physical limits of the jet flow along the $x$-axis and the optical depth of the jet material is given by
\begin{equation}
\tau (l)= \chi_{\nu} r(x) = \chi _0 r_0 \nu^{\left(-p-4\right)/2} l^{a_1-a_3-a_2 \left(p+2 \right)/2}.
\label{depth}
\end{equation}
The reduction in the number of Doppler factors arises from our assumption of a steady state of the jet flow. In principle further relativistic corrections must be applied in the case of mixed optically thin and thick emission, but these corrections are small \citep{tc91} and we neglect them here for simplicity.

It is convenient to recast equation (\ref{lflux}) with the help of equation (\ref{depth}) as an integration over optical depth,
\begin{equation}
F_{\nu} = \delta _{\mp}^2 \frac{x_0 r_0 J_0}{a_4 D^2 \chi _0} \nu^{5/2} \tau _0^{-a_5} \int _{\tau _{\rm max}}^{\tau _{\rm min}} \tau^{a_5-1} \left( 1- e^{-\tau} \right) \, {\rm d} \tau,
\label{flux}
\end{equation}
where 
\begin{equation}
a_4 = 2a_1 -2a_3-\left(p+2\right) a_2
\end{equation}
and
\begin{equation}
a_5 = \frac{2a_1+a_2+2}{a_4}.
\end{equation}
The coefficients $a_4$ and $a_5$ are listed for the ballistic and adiabatic jets in Table \ref{expo}. $\tau_0$ is given by setting $l=1$ in equation (\ref{depth}) while $\tau_{\rm max} = \tau (l_{\rm min})$ and $\tau_{\rm min} = \tau (l_{\rm max})$, which reflects the fact that the optical depth is always greatest in the innermost regions of the jet. The solution of equation (\ref{flux}) is given by
\begin{equation}
F_{\nu} = \delta _{\mp}^2 \frac{x_0 r_0 J_0}{a_4 D^2 \chi_0} \nu^{5/2} \tau_0^{-a_5} \left[ \Gamma \left(a_5, \tau \right)+\frac{1}{a_5} \tau^{a_5} \right]_{\tau_{\rm max}}^{\tau_{\rm min}}.
\label{sol}
\end{equation}
Here, the incomplete $\Gamma$-function is defined as 
\begin{equation}
\Gamma \left( a , z \right) = \int_z^{\infty} t^{a-1} e^{-t} \, {\rm d} t.
\end{equation}

\subsection{Spectra in the absence of radiative energy losses and without a high-energy cut-off}
\label{simple}

We can immediately recover the well-known solutions for an entirely optically thin ($\tau _{\rm min} \ll 1$ and $\tau _{\rm max} \ll 1$) and an entirely optically thick ($\tau_{\rm min} \gg 1$ and $\tau_{\rm max} \gg 1$) jet. We note that for all choices of $a_1$ discussed above and for physical reasonable values for the power-law exponent $2\le p \le 3$ we find $a_5 < 0$ (see Table \ref{expo}). The incomplete $\Gamma$-function has the series representation \citep{gr00}
\begin{equation}
\Gamma \left(a, z \right) = \Gamma \left( a \right) - \sum_{n=0}^{\infty} \frac{\left( -1 \right)^n z^{a+n}}{n! \left( a+n \right)}.
\label{series}
\end{equation}
For $\tau \ll 1$ we ignore all terms beyond $n=1$ and thus obtain
\begin{equation}
\left[ \Gamma \left( a_5, \tau \right)+\frac{1}{a_5} \tau^{a_5} \right]_{\tau_{\rm max}}^{\tau_{\rm min}} \sim \frac{1}{1+a_5} \left( \tau _{\rm min}^{1+a_5} - \tau _{\rm max}^{1+a_5} \right).
\end{equation}
Because of equation (\ref{depth}) we have $\tau_0 \propto \tau_{\rm min} \propto \tau _{\rm max} \propto \nu^{\left(-4-p\right)/2}$, and from equation (\ref{sol}) it then follows that $F_{\nu} \propto \nu^{\left(1-p \right)/2}$ as expected.

For large optical depths we can use the limit for the incomplete $\Gamma$-function, $\lim_{z\rightarrow \infty} \Gamma \left(a, z \right) =0$ \citep{gr00}. Thus
\begin{equation}
\left[ \Gamma \left( a_5, \tau \right)+\frac{1}{a_5} \tau^{a_5} \right]_{\tau_{\rm max}}^{\tau_{\rm min}} \sim \frac{1}{a_5} \left( \tau_{\rm min}^{a_5} - \tau_{\rm max}^{a_5} \right),
\label{largetau}
\end{equation}
and from the proportionality of the optical depths terms it then follows that $F_{\nu} \propto \nu^{5/2}$, again as expected.

\begin{figure}
\centerline{
\includegraphics[width=8.45cm]{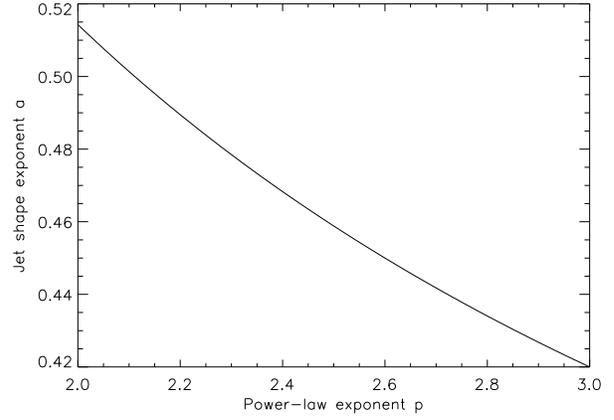}}
\caption{Relation between the geometrical shape of the jet of model A3 and the power-law exponent of the energy distribution of the relativistic electrons for which the jet emission spectrum is flat. The plot shows the relation in equation (\ref{flatrel}).}
\label{flat}
\end{figure}

The final special case is that of a spatially very extended jet or `long' jet. If the physical dimensions of the long jet, $l_{\rm min}$ and $l_{\rm max}$, are such that $\tau_{\rm max} \rightarrow \infty$ and $\tau_{\rm min} \rightarrow 0$, then for $-1 < a_5 < 0$ we have
\begin{equation}
\left[ \Gamma \left( a_5, \tau \right)+\frac{1}{a_5} \tau^{a_5} \right]_{\tau_{\rm max}}^{\tau_{\rm min}} \sim \Gamma \left( a_5 \right),
\end{equation}
which implies
\begin{equation}
F_{\nu} \propto \nu^{\left[5 +\left( p+4 \right) a_5\right]/2}.
\label{longslope}
\end{equation}
Using the results summarised in Table \ref{expo}, we recover the result of \citet{bk79} that the ballistic jet with the magnetic field perpendicular to the jet axis (model B1) has a flat spectrum, i.e. $F_{\nu}$ is independent of $\nu$, if it is extended and $-1 < a_5 < 0$. The ballistic jet with a parallel magnetic field, model B2, can never produce a flat spectrum for positive $p$ because equation (\ref{longslope}) predicts $F_{\nu} \propto \nu^ {\left(7p +3 \right) / \left[2 \left( 2p + 3\right) \right]}$. For the adiabatic jet models we can substitute the expressions for $a_5$ and find that a flat spectrum is predicted if the exponent $a$ for the geometrical shape of the jet, $r(x) \propto x^a$, is given by
\begin{eqnarray}
{\rm Model\ A1:} && a = \frac{3p +12}{13p +2}\nonumber\\
{\rm Model\ A2:} && a = \frac{3p+12}{19p+11}\\
{\rm Model\ A3:} && a = \frac{3p+12}{15p+5} \nonumber.
\label{flatrel}
\end{eqnarray}
Figure \ref{flat} plots the relation for model A3. It is interesting that for all adiabatic jets geometrical shapes described by exponents $a$ in the range $1/3 \ltappeq a \ltappeq 2/3$ are required for flat spectra.

Note that for $a_5 < -1$ the $n=1$ term in the series in equation (\ref{series}) dominates for $\tau_{\rm min} \rightarrow 0$. In that case we have for the long jet
\begin{equation}
\left[ \Gamma \left( a_5, \tau \right)+\frac{1}{a_5} \tau^{a_5} \right]_{\tau_{\rm max}}^{\tau_{\rm min}} \sim \frac{1}{1+a_5} \tau _{\rm min}^{1+a_5},
\end{equation}
similar to the entirely optically thin jet. The spectrum of the long jet is then also optically thin, i.e. $F_{\nu} \propto \nu^{\left(1-p \right)/2}$, and a flat spectrum is not possible for physically reasonable values of the exponent $p$. 

\begin{figure}
\centerline{
\includegraphics[width=8.45cm]{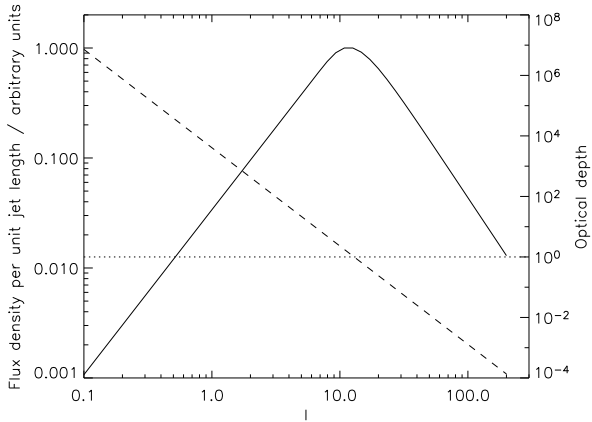}}
\caption{Flux density per unit length (solid line) at a single frequency as a function of position along the jet axis for the ballistic jet. This plots the integrand in equation (\ref{lflux}). The optical depth at the same frequency is also given (dashed line).}
\label{tauspec}
\end{figure}

\begin{figure}
\centerline{
\includegraphics[width=8.45cm]{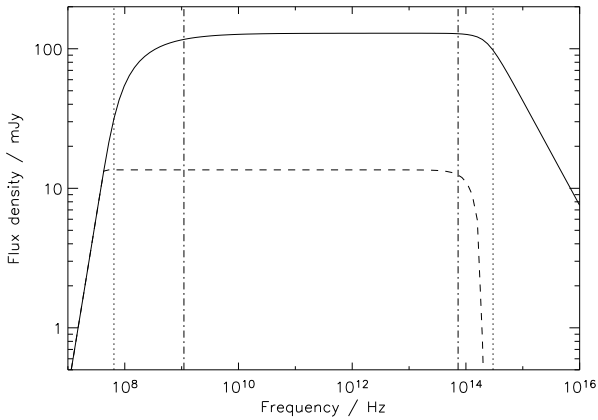}}
\caption{Spectrum of the ballistic jet model B1. The solid line shows the spectrum without radiative losses or high-energy cut-off in the electron energy density distribution. The short-dashed line is the same spectrum including synchrotron losses and a high-energy cut-off. The dotted lines show the frequency at which the optical depth at the beginning (high frequency) and the end (low frequency) of the jet, $l_{\rm min}$ and $l_{\rm max}$ respectively, is equal to unity. The dot-dashed line at low frequencies indicates where the optical depth is equal to $10^{-4}$ at $l_{\rm max}$. The dot-dashed line at high frequency shows where the optical depth is 100 at $l_{\rm min}$.}
\label{illustration2}
\end{figure}

\begin{table}
\begin{tabular}{lc}
Model parameter & Value\\
\hline
$x_0$ & $47$\,AU\\
$r_0$ & $8.9 \times 10^7$\,m\\[1ex]
$p$ & $2.5$\\
$B_0$ & $2.4$\,mT\\[1ex]
$\kappa _0$ & $3.3\times10^{-7}$\,J$^{1.5}$\,m$^{-3}$\\
$D$ & $2$\,kpc\\[1ex]
$l_{\rm min}$ & $4.3\times 10^{-5}$\\
$l_{\rm max}$ & $200$\\[1ex]
$\gamma_{\rm max} \left( t_{\rm min} \right)$ & $10^6$\\
$v_{\rm j}$ & $0.97\,c$\\
$\vartheta$ & $40^{\circ}$\\
\hline
\end{tabular}
\caption{Parameters used to illustrate the spectral properties of the jet models. These model parameters are also used to explain the observational data of Cygnus X-1 in Section \ref{cyg} when using a ballistic jet model with a magnetic field perpendicular to the jet axis, model B1. \label{modpara}}
\end{table}

\begin{figure}
\centerline{
\includegraphics[width=8.45cm]{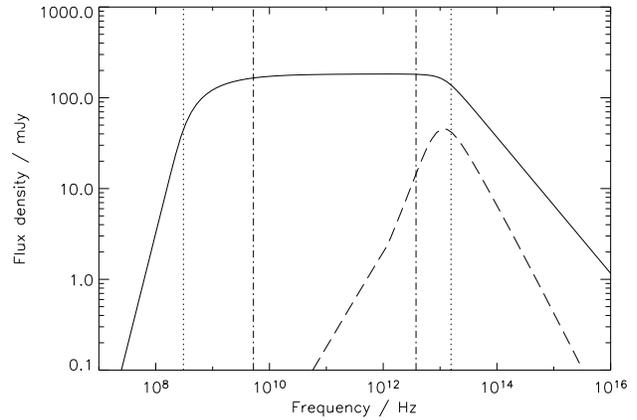}}
\caption{Same as Figure \ref{illustration2}, but for the adiabatic jet model A3 with $a=0.46$. The long-dashed line shows the spectrum including both adiabatic and synchrotron energy losses.}
\label{adillu2}
\end{figure}

Unless the jet is exceedingly short, there will always be a range of frequencies for which $\tau_{\rm min} \rightarrow 0$ and $\tau_{\rm max} \rightarrow \infty$ and the long jet scenario applies. An example is shown in Figure \ref{tauspec} where we plot the integrand in equation (\ref{lflux}) and the optical depth of the jet material as a function of $l$ for a single frequency for the ballistic jet. The contribution to the overall flux of the jet peaks close to $\tau=1$. For the long jet scenario to apply the jet must be long enough so that substantial emission from either side of the peak contributes to the overall flux. 

Below the frequency range of the long jet the spectrum will follow the optically thick case, $F_{\nu} \propto \nu^{5/2}$, and above this range the optically thin case applies with $F_{\nu} \propto \nu^{(1-p)/2}$. The solid line in Figure \ref{illustration2} illustrates this generic overall shape of the jet spectrum for the ballistic jet model B1. For this Figure and the following we have used the model parameters summarised in Table \ref{modpara}. The observations of the partially self-absorbed jet of Cyg X-1 are well explained by the model for these parameters (see Sections \ref{losses} and \ref{cyg}). For comparison, the solid line in Figure \ref{adillu2} shows the spectrum of the adiabatic jet model A3 with $a=0.46$ for the same set of parameters. The value of $a$ was chosen according to equation (\ref{longslope}) to allow for a flat spectrum at intermediate frequencies.

It is interesting to note that the frequencies for which the optical depth of the jet material is unity at $l_{\rm min}$ and $l_{\rm max}$ are located well within the optically thick and thin regimes respectively. The transition from the, in this case, flat spectrum of the long jet occurs closer to $\tau \left( l_{\rm max} \right) \sim 10^{-4}$ and $\tau \left( l_{\rm min} \right) \sim 100$. Obviously, for sufficiently short jets the frequency range over which the long jet case applies may vanish altogether. 

\section{Including energy losses of the electrons}

In the previous Section we did not consider the effect of radiative energy losses of the relativistic electrons on the predicted spectra. The adiabatic jet models include the effect of adiabatic energy losses on the overall energy distribution of the relativistic electrons. However, because we did not impose a high-energy cut-off to this distribution, we did not have to consider the effect of adiabatic losses on such a cut-off. In this Section we introduce a high-energy cut-off at $\gamma _{\rm max}$ and include the effect of adiabatic energy losses on this cut-off. 

\subsection{Evolution of the high-energy cut-off}
\label{cutoff}

\subsubsection{Adiabatic and synchrotron losses}

Other than adiabatic energy losses, the radiative losses due to synchrotron radiation modify the energy distribution of the relativistic electrons away from a simple power-law, unless $p=2$ \citep[e.g.][]{nk62}. In the following we will make the simplifying assumption that the energy losses only shift the sharp high-energy cut-off to lower energies while not altering the power-law shape or exponent of the power-law distribution. This approximation does not introduce a large error as the deviation from the original power-law is significant only near the cut-off. Also, the expressions for the synchrotron emissivity and absorption coefficient given in equations (\ref{jk}) and (\ref{jk0}) are strictly valid only for power-law energy distribution extending from $\gamma =1$ to $\gamma_{\rm max} \rightarrow \infty$. However, the expressions involved in the derivation of $J_{\nu}$ and $\chi_{\nu}$ decay sufficiently quickly for $\gamma \gtappeq 10$ that the results for finite $\gamma _{\rm max}$ do not deviate greatly from those presented in the previous Section for $\gamma_{\rm max} \rightarrow \infty$ \citep{rl79,ml94}.

In the optically thin regime the evolution of the Lorentz factor of a given relativistic electron in the rest frame of the jet material is described by \citep[e.g.][]{ml94}
\begin{equation}
\dot{\gamma} = -\frac{4}{3} \frac{\sigma _{\rm T} u_0}{m_{\rm e} c} \left( \frac{t}{t_0} \right)^{-2 a_2} \gamma^2 - \frac{2a_1}{3t} \gamma,
\label{evol}
\end{equation}
where the first term on the right describes the energy losses due to synchrotron radiation and the second term reproduces equation (\ref{adiabat}) for the adiabatic losses where we substituted for $\Delta V$. Because of our assumption of a constant bulk velocity for the jet material along the $x$-axis, $v_{\rm j}$, we can express the dimensionless coordinate $l=x/x_0$ also as a time variable, i.e. $l=\gamma _{\rm j}v_{\rm j}t/x_0$. Because of time dilation, we have to include $\gamma_{\rm j}$. Thus $t=l x_0 / \left( v_{\rm j} \gamma _{\rm j} \right)$ and $t_0=x_0/ \left( v_{\rm j} \gamma _{\rm j} \right)$. $\sigma _{\rm T}$ is the Thomson cross section and $u_0=B_0^2/\left(2 \mu_0 \right)$ is the energy density of the magnetic field at $x_0$. The solution of equation (\ref{evol}) is found as
\begin{equation}
\gamma \left( t \right) = \frac{\gamma \left( t_{\rm min}\right) t^{-2 a_1 /3}}{t_{\rm min}^{-2 a_1/3} + \frac{4 \sigma _{\rm T} u_0}{3 a_6 m_{\rm e} c} t_0^{2a_2} \gamma \left( t_{\rm min}\right) \left( t^{a_6}-t_{\rm min}^{a_6} \right)},
\label{gamadiabat}
\end{equation}
with $a_6 = 1-2a_2-2a_1/3$ and $t_{\rm min} = l_{\rm min} x_0 / \left( v_{\rm j} \gamma _{\rm j} \right)$. Electrons which were injected into the jet at $t_{\rm min}$ or, equivalently, $l_{\rm min}$ with a Lorentz factor $\gamma \left( t_{\rm min}\right)$, have a Lorentz factor $\gamma \left( t \right)$ at $t$ or, equivalently, $l$. For the ballistic jet the adiabatic, second term on the right of equation (\ref{evol}) vanishes and we have instead
\begin{equation}
\gamma \left( t \right) = \frac{\gamma \left( t_{\rm min}\right)}{1+\frac{4 \sigma _{\rm T} u_0}{3 \left(1-2a_2 \right) m_{\rm e} c} t_0^{2a_2} \gamma \left( t_{\rm min} \right)\left( t^{1-2a_2}-t_{\rm min}^{1-2a_2} \right)}.
\label{gamrad}
\end{equation}

Note that the exponents $a_6$ and $1-2a_2$ are usually negative. This implies in the case of the ballistic jet models (no adiabatic losses) that the Lorentz factors of electrons do not necessarily decrease forever, but converge to a finite value for $t\gg t_{\rm min}$. The somewhat surprising result simply reflects the fact that the synchrotron losses rapidly decline in the decreasing magnetic field of the expanding jet. For the adiabatic case the adiabatic losses continue at all times and so $\gamma \propto t^{-2a_1 /3} \propto l^{-2 a_1 / 3}$ for $t\gg t_{\rm min}$. 

For optically thin conditions the evolution of the high-energy cut-off $\gamma _{\rm max}$ also obeys equations (\ref{gamadiabat}) and (\ref{gamrad}). Below we will refer to the high-energy cut-off in the optically thin regime as $\gamma _{\rm thin}$. However, for large parts of the spectrum the jet is optically thick. Electrons with a given Lorentz factor $\gamma$ emit most of their radiation at the critical frequency $\nu \sim \nu _{\rm g} \gamma^2$, where the gyro-frequency is defined as $\nu_{\rm g} = e B / \left(2 \pi m_{\rm e}\right)$. An electron emitting at a critical frequency for which the jet is optically thick gains energy through synchrotron self-absorption. Ideally we would include an energy gain term for the self-absorption effect into equation (\ref{evol}) and then derive the electron evolution as before. While this approach leads to analytic solutions when only considering the systematic energy gain of electrons of a single energy \citep{mr67}, it is not applicable in most cases because the stochastic energy gain for power-law energy distributions is comparable to the systematic term. In this case, only numerical solutions are possible, because the stochastic term depends on the entire energy distribution \citep{rm69}. 

The full numerical treatment of synchrotron losses and gains in the optically thick regime is beyond the scope of this paper. However, electrons radiating mainly at frequencies for which the jet is optically thick, do on average not lose or gain energy due to radiative effects, even if they are relatively close to the surface of the jet \citep{rm69}. Thus the high-energy cut-off in the optically thick regime, $\gamma _{\rm thick}$, is given by the requirement that $\tau \left( \gamma _{\rm thick} \right) \sim 1$. The electrons at this cut-off emit mainly at a frequency $\nu_{\rm thick} = \nu_{\rm g} \gamma _{\rm thick}^2$ and so we find from equation (\ref{depth})
\begin{equation}
\gamma _{\rm thick} = \left[ \frac{4 \pi m_{\rm e}}{3 e B_0} \left( \chi_0 r_0 \right)^{2 / \left(p+4 \right)} l^{a_7} \right]^{1/2},
\label{gamthick}
\end{equation}
with
\begin{equation}
a_7 = \left(a_1 -a_3 - a_2 \frac{p+2}{2}\right) \frac{2}{p+4}+a_2.
\end{equation}
For the ballistic jet with perpendicular magnetic field, model B1, $a_7 =0$ and so $\gamma _{\rm thick}$ is constant along the entire length of the jet. In other words, electrons with Lorentz factors equal or below $\gamma _{\rm thick}$ never loose their energy to radiation unless they are very close to the jet surface. The existence of a constant high-energy cut-off is required for a flat spectrum from the jet. While \citet{bk79} invoked an unknown re-acceleration mechanism to ensure $\gamma _{\rm max} ={\rm constant}$, we have shown here that such a process is unnecessary because of the energy gains associated with synchrotron self-absorption. For a parallel magnetic field in the ballistic jet, model B2, we have $a_7= 2/ \left(p+4 \right)$. The Lorentz factor of relativistic electrons for which the jet is optically thick is {\em increasing\/} for increasing $l$ in this model. Therefore, if $l_{\rm thick}$ is the position along the jet where $\gamma _{\rm thin} = \gamma _{\rm thick}$, we have $\gamma_{\rm max} =\gamma_{\rm thick}$ at this position and $\gamma _{\rm max} = {\rm constant}$ for all $l>l_{\rm thick}$. In the adiabatic cases we find
\begin{eqnarray}
{\rm Model\ A1:} && a_7 = a \frac{4 \left( 1 - p \right)}{3 \left( p + 4 \right)}\nonumber\\
{\rm Model\ A2:} && a_7 = a \frac{2 \left( 5 - 2p \right)}{3 \left( p + 4 \right)}\\
{\rm Model\ A3:} && a_7 = a \frac{2 \left( 3 -2 p \right)}{3 \left(p+4 \right)}.\nonumber
\end{eqnarray}

Even for optically thick conditions the electron energy distribution does not deviate greatly from the original power-law with a high-energy cut-off $\gamma _{\rm max}$ for exponents $2\le p \le 3$ \citep{rm69}. Thus, for each position $l$ along the jet we can now determine $\gamma_{\rm max}$ and thereby the entire electron energy distribution. For small $l$ the cut-off is given by $\gamma _{\rm max} =\gamma _{\rm thin}$. Further down the jet $\gamma _{\rm thin}$ will first become equal to and then fall below $\gamma _{\rm thick}$ and for the ballistic jet models $\gamma _{\rm max} = \gamma _{\rm thick}$ afterwards. In the adiabatic jet models two competing effects can diminish $\gamma _{\rm max}$ further after passing through the point at which $\gamma _{\rm thick} = \gamma _{\rm thin}$. In most cases equation (\ref{gamthick}) implies a further reduction of $\gamma _{\rm thick}$ for increasing $l$. This means that $\gamma _{\rm max}$ would also decrease. At the same time, adiabatic losses lead to $\gamma _{\rm max} = \gamma _{\rm thick} \left( l / l_{\rm thick} \right) ^{-2 a_1 / 3}$. It is straightforward to show that for all adiabatic models the adiabatic losses of $\gamma _{\rm max}$ are the dominant effect. 

By again making the assumption that all electrons only emit at their critical frequency, we can now define for a given frequency $\nu$ a maximum distance $l'_{\rm max}$ along the jet axis where the jet material is still contributing to the overall emission,
\begin{equation}
l'_{\rm max} = \left(\frac{2 \pi m_{\rm e} \nu}{e B_0 \gamma _{\rm max}^2} \right)^{-1/a_2}.
\label{ldmax}
\end{equation}
Clearly, as long as $l'_{\rm max}$ is larger than the physical extent of the jet, $l_{\rm max}$, the jet spectrum is not affected by energy losses of the relativistic electrons at frequency $\nu$ and we can use the results of the previous Section. For $l'_{\rm max} < l_{\rm max}$ we have to take into account energy losses of the electrons by using $l'_{\rm max}$ instead of $l_{\rm max}$ in the calculation of $\tau_{\rm min}$.

\subsubsection{Losses due to Compton scattering}

In the optical thick parts of the jet Compton scattering of the synchrotron photons off the relativistic electrons may become important. We do not include energy losses of the relativistic electrons due to Compton scattering in the jet in our calculations as these would require a full treatment of radiative transfer. However, it is obviously necessary to test whether these losses are important when applying the model to observational data and so we give the necessary expressions below. 

The relevant limit for the energy density of the synchrotron photon field is most conveniently expressed in terms of the brightness temperature \citep[e.g.][]{rl79},
\begin{equation}
T_{\rm b} = \frac{c^2 I_{\nu}}{2 \delta_{\mp}^3 k_{\rm B} \nu^2},
\end{equation}
where $k_{\rm B}$ is the Boltzman constant and $\nu$ is the emitted frequency rather than the observing frequency. For $T_{\rm b} \ltappeq 10^{12}$\,K Compton losses are not important compared to the energy losses due to synchrotron radiation. The maximum brightness temperature for a given frequency is reached at the position along the jet where the optical depth of the jet material roughly equals unity for photons of this frequency. Hence in our model we have
\begin{equation}
T_{\rm b, max} = \frac{c^2 J_0}{8 \pi k_{\rm B} \chi _0} \left( \chi _0 r_0 \right)^{1/ \left( p+4 \right)} \left( 1 - e^{-1} \right) l^{ \left( a_1 + a_2 - a_3 \right) / \left( p + 4 \right)}.
\label{bright}
\end{equation}
The maximum brightness temperature is either constant or only a weak function of $l$ in all our models. Also, using the dependencies of $J_0$ and $\chi _0$ given in equation (\ref{jk0}) we find that $T_{\rm b, max}$ depends only weakly on the other model parameters $B_0$, $\kappa _0$ and $r_0$. 

In some jets relativistic induced Compton scattering may be more important than direct Compton scattering discussed above. Induced Compton scattering causes significant energy losses for the relativistic electrons if \citep{sk94}
\begin{equation}
\frac{k_{\rm B} T_{\rm b}}{m_{\rm e} c^2} \tau _{\rm T} \ge 1,
\label{crit}
\end{equation}
where $\tau_{\rm T}$ is the Thomson depth of the jet material,
\begin{equation}
\tau_{\rm T} = n_{\rm e} \sigma_{\rm T} r.
\end{equation}
Here, $n_{\rm e}$ is the number density of electrons, $\sigma _{\rm T}$ is the Thomson cross-section and we have again assumed that the average path length a photon travels through the jet material is equal to the jet radius, $r$. For our power-law energy distribution of the electrons an upper limit for the electron density is given by $n_{\rm e} \le \kappa \left( m_{\rm e} c^2 \right)^{1-p}$. The Thomson depth of the jet materials in our models is then limited by
\begin{equation}
\tau_{\rm T} \le \left(m _{\rm e} c^2 \right)^{1-p} \sigma _{\rm T} \kappa _0 r_0 l^{a_1-a_3}.
\label{thom}
\end{equation}

Equations (\ref{bright}) and (\ref{thom}) can be used to ensure that the models are applicable to a given observational data set, i.e. that they do not suffer from Compton losses which are not included in the models. 

\subsection{Spectra with energy losses and a high-energy cut-off}
\label{losses}

\subsubsection{Iterative construction of model spectra}

For a given set of model parameters we can construct a model spectrum. In practice this will involve the determination of $\tau_{\rm min}$ and $\tau _{\rm max}$ from equation (\ref{depth}) to be substituted into equation (\ref{sol}). For a given frequency $\nu$, we set $l= l_{\rm min}$ and calculate $\tau_{\rm max}$. Determining $\tau _{\rm min}$ is more involved as it requires the calculation of $l'_{\rm max}$. This calculation involves an implicit equation and so cannot be done analytically. Here we describe one possible iterative procedure for the determination of $l'_{\rm max}$. 

The first step is to choose a trial distance $l$ such that $l_{\rm min} \le l \le l_{\rm max}$.  The strength of the magnetic field in the jet material at $l$ is $B \left( l \right) = B_0 l^{-a_2}$. The maximum frequency at which jet material located at $l$ is still contributing to the emission is given by $\nu _{\rm max} \left( l \right) = \nu_{\rm g} \left( l \right) \gamma _{\rm max} \left( l \right)$. For the next iteration we need to compare $\nu_{\rm max} \left( l \right)$  with $\nu$. Therefore we must next derive $\gamma _{\rm max} \left( l \right)$, the high-energy cut-off of the electron energy distribution at $l$. 

From equations (\ref{gamadiabat}, adiabatic jet models) or (\ref{gamrad}, ballistic jet models) we can determine $\gamma _{\rm thin} \left( l \right)$. We calculate $\gamma _{\rm thick}$ from equation (\ref{gamthick}). If $\gamma_{\rm thin} \left( l \right) > \gamma _{\rm thick}$, then $\gamma _{\rm max} \left( l \right)= \gamma _{\rm thin} \left( l \right)$. Otherwise, for the ballistic jet models $\gamma _{\rm max} \left( l \right) = \gamma _{\rm thick}$. For the adiabatic jet models $\gamma _{\rm max} \left( l \right) = \gamma _{\rm thick} \left( l / l_{\rm thick} \right)^{-2a_1 /3}$. The required distance $l_{\rm thick}$ must be found from the implicit equation resulting from setting $\gamma \left( t \right) = \gamma _{\rm thick}$ in equation (\ref{gamadiabat}). Finally, if $\nu_{\rm max} \left( l \right) > \nu$, then the trial distance in the next iteration should be larger than the current one. In the case of $\nu_{\rm max} \left( l \right) < \nu$, the trial distance should be decreased. The iterations can be stopped when $\nu_{\rm max} \left( l \right) \sim \nu$ within the required accuracy and at that point we can set $l'_{\rm max} =l$ and then proceed to calculate $\tau_{\rm min}$.

\subsubsection{Example spectra}

The model parameters in Table \ref{modpara} were chosen to explain the observations of the jet in Cygnus X-1 with the ballistic jet model B1 including radiative energy losses (see Section \ref{cyg}). This does not imply that energy losses will always be important in all jets and at all frequencies. 

The spectrum of the ballistic and adiabatic jets with energy losses of the electrons are shown in Figures \ref{illustration2} for model B1 and \ref{adillu2} for model A3. When energy losses of the electrons are taken into account, then we cannot in general expect that $\tau_{\rm min} \ll 1$ for a given frequency. Therefore the spectrum of the jet will not necessarily be that of the long jet described by equation (\ref{longslope}). In fact, in most cases $\tau_{\rm min}$ will considerably exceed unity. If $\tau_{\rm max} \rightarrow \infty$, then equation (\ref{largetau}) applies and we find
\begin{equation}
F_{\nu} \sim  \delta_{\mp}^2 \frac{x_0 r_0 J_0}{a_4 a_5 D^2 \chi_0} \nu^{5/2} \left( \frac{\tau_{\rm min}}{\tau_0} \right)^{a_5}.
\label{reduced}
\end{equation}

For the ballistic jet models we have argued in the previous Section that $\gamma _{\rm max} ={\rm constant}$ for $l \ge l_{\rm thick}$. For rapid radiative energy losses of the electrons at $l$ in the range $l_{\rm min} < l < l_{\rm thick}$, the distance $l_{\rm thick}$ itself will not depend strongly on the observing frequency. From equation (\ref{ldmax}) we then find $l'_{\rm max} \propto \nu ^{-1/a_2}$ and substituting into equation (\ref{depth}) we get
\begin{equation}
\tau_{\rm min} \propto \nu^{\left( a_3 -a_1-a_2 \right) / a_2}.
\label{taumin}
\end{equation}
Note that the exponent does not depend on the slope of the electron energy distribution, $p$, nor on the geometrical shape of the jet described by $a$ (see Table \ref{expo}). Finally, from equation (\ref{reduced}) we obtain the slope of the spectrum as
\begin{equation}
F_{\nu} \propto \nu^{\left( 4a_2-2a_1-2 \right) / \left( 2 a_2 \right)}.
\end{equation}
For the ballistic jet with perpendicular magnetic field, model B1, we have $F_{\nu} ={\rm constant}$ as in the case without energy losses of the electrons, which is confirmed by Figure \ref{illustration2}. For a parallel field structure, model B2, the spectrum would follow $F_{\nu} \propto \nu$. Model B2 is incompatible with a flat spectrum. The slope in the optically thin part of the spectrum in Figure \ref{illustration2} is steeper compared to the case of no energy losses because of the effect of the high-energy cut-off in the electron energy spectrum.

In the case of the adiabatic jet models $\gamma _{\rm max} \propto l^{-2 a_1 /3}$ and so $l'_{\rm max} \propto \nu^{-3 / \left( 4 a_1 + 3 a_2 \right)}$. Again substituting into equation (\ref{depth}) yields
\begin{equation}
\tau_{\rm min} \propto \nu^{- \left(7 a_1 + 3 a_2 \right) / \left( 4 a_1 + 3 a_2 \right)},
\end{equation}
where we also used $a_3 = \left( 4 + 2p \right) a_1 /3$ as appropriate for the adiabatic jet models. Again the exponent of this expression does not depend on $p$ or $a_1$. The shape of the spectrum is now
\begin{equation}
F_{\nu} \propto \nu^{\left( 7 a_1 + 6 a_2 -3 \right) / \left(4 a_1 +3 a_2 \right)}.
\end{equation}
The exponent of the power-law spectrum predicted by model A3 for $a_1 = 0.46$ is then 1.06 which is confirmed by the slope of the spectrum in Figure \ref{adillu2} below about $10^{12}$\,Hz. The emission at high frequencies comes from the innermost parts of the jet close to $l_{\rm min}$. Radiative losses had not enough time there to completely change the electron energy distribution. This explains the peak in the spectrum. At the highest frequencies only optically thin parts of the jet contribute to the overall emission and lead to a negative power-law similar to the case without energy losses. Note however that the slope of this power-law is somewhat steeper due to the decreasing high-energy cut-off. 

The adiabatic jet models are all consistent with a flat spectrum provided the shape of the jet described by $a_1$ takes a suitable value (Model A1: $a_1 = 3/13$, A2: $a_1 = 3/19$ and A3: $a_1 = 1/5$). In all three adiabatic models the jet needs to be strongly confined, i.e. $a_1 < 1/4$, to achieve a flat emission spectrum. The slope of the spectrum is quite sensitive to the value of $a_1$. For example, a change of $a_1$ in model A3 from $1/5$ to $1/3$ results in a change of the power law exponent of the spectrum from zero to 0.75.

For the following discussion we note that all the spectral slopes calculated above, with the only exception of the purely optically thin case, are all independent of the exponent $p$ of the power-law describing the energy distribution of the electrons. Therefore we can readily apply the model to observational data even if the optically thin part of the spectrum is not observed and thus we do not know the value of $p$.

\section{Application to observations}
\label{obs}

The jet emission models depend on a number of parameters. Some of these parameters can be constrained by applying general considerations and others may be inferred from applying the model predictions to observational data with a view to determining the physical conditions within the jet. Here we discuss all of the relevant parameters in turn and demonstrate below how observations of the jet in Cygnus X-1 may be used to infer the properties of this object.

The scale height $x_0$ can always be chosen arbitrarily to provide a convenient location along the jet axis at which to define the exact values of other quantities. In many cases we will be mainly interested in that part of the jet spectrum which is strongly affected by absorption. As we have seen in the previous Section, we then do not need to know the exponent of the power-law energy distribution of the electrons, $p$, as it does not influence the slope of the predicted spectrum. However, estimates for other quantities derived from the model, for example the strength of the magnetic field, depend weakly on $p$. The distance of the jet, $D$, the bulk velocity of the jet material, $v_{\rm j}$, and the viewing angle of the jet axis to our line of sight, $\vartheta$, cannot normally be determined by the model itself and need to be measured by other means. 

The parameters describing the geometrical shape and size of the jet, $r_0$, $l_{\rm min}$ and $l_{\rm max}$, could in principle be determined from observations. However, $l_{\rm min}$ is probably too small to be resolvable even with a large improvement on current resolution limits. Currently only upper limits exist for the radius of jets in Galactic X-ray binaries  (Miller-Jones et al., in preparation) while for AGN jets $r_0$ is sometimes resolved \citep[e.g.][]{jbl99}. The maximum extent of a jet at a given observing frequency is sometimes measured and an example is provided by the observations of the jet in Cygnus X-1 of \citet{ssf01} which we use in the following Section. It should be borne in mind that at one observing frequency  we can always only measure $l'_{\rm max}$ given by equation (\ref{ldmax}) rather than the physical extent of the jet flow $l_{\rm max}$. However, $l_{\rm max}$ only determines the low frequency cut-off of the spectrum, but is not important for the model otherwise. In the case of the adiabatic jet models resolved observations can, in principle, also determine the shape of the jet as described by the parameter $a$. However, in practice it is easier to infer the value of $a$ from the slope of the observed self-absorbed spectrum as this is a strong function of $a$. 

Finally, the normalization of the electron energy distribution, $\kappa _0$, and the strength of the magnetic field, $B_0$, cannot be determined directly from observations, but must be inferred from the model. We can reduce the number of free parameters by assuming that the energy densities of the magnetic field and of the relativistic electrons are initially in equipartition. In this case,
\begin{equation}
u_0 = \frac{B_0^2}{2 \mu _0} = \int_{E_{\rm min}}^{E_{\rm max}} \kappa _0 E^{1-p} \, {\rm d}E.
\end{equation}
For an energy distribution extending over all physically meaningful Lorentz factors ($1 \le \gamma \le \infty$) we then have
\begin{equation}
\kappa _0 \sim \left( p-2 \right) \frac{B_0^2}{2 \mu _0} \left( m_{\rm e} c^2 \right)^{p-2}.
\end{equation}

With these considerations we can now apply the model to observations and determine relevant parameters for the observed jet.

\subsection{Application to Cygnus X-1}
\label{cyg}

In the following we apply the model to the jet observed in the X-ray binary Cygnus X-1. \citet{ssf01} report a resolved jet extending to about 15\,mas from the position of the X-ray binary system at an observing frequency of 8.4\,GHz. We set the bulk velocity of the jet material to 0.97\,$c$ and the viewing angle to our line of sight to $\vartheta = 40^{\circ}$ \citep{ssf01}. For a distance of 2\,kpc \citep{gzp99} the observed, projected jet length then corresponds to a real jet length of roughly 47\,AU. For convenience we set $x_0$ equal to this value and so $l'_{\rm max} = 1$ at 8.4\,GHz. Only one jet is observed and it is therefore reasonable to assume that this is the approaching jet. Note that the bulk velocity of the jet and the viewing angle imply $\delta _- \sim 1$ and so $\nu \sim \nu_{\rm ob}$. The total flux density at the same frequency is 13\,mJy. 

In a map at 15\,GHz from an earlier observing epoch, the jet may also be marginally extended along the same axis \citep{ssg98,ssf01}. The extension is 2\,mas or less which corresponds to $l'_{\rm max} \left( 15\,{\rm GHz} \right) \le 0.13$. In the following we will mainly concentrate on the observational data at 8.4\,GHz.

There are no simultaneous observations at any frequency other than 8.4\,GHz and so we cannot be certain what the spectral slope of the jet emission was. However, during the low/hard X-ray state the source usually shows a flat spectrum extending up to at least 220\,GHz \citep{fpdtb99}. We assume here that the spectral slope at the time the radio jet at 8.4\,GHz was observed was zero. The extent of the flat spectrum to high frequencies is also not known. However, for GX 339-4 the flat jet spectrum is observed to near-IR wavelengths \citep{cf02} while for XTE J1118+480 it may extend to the near-UV \citep{hmh00}. For the purpose of illustrating the model, we assume here that the flat spectrum extends to the near-IR of wavelengths of about 1\,$\mu$m. The flat spectrum always arises from regions in the jet which are at least partially self-absorbed. Thus the exact value of $p$ is not very important and we set $p=2.5$. 

\subsubsection{Ballistic jet models}

We have seen above that the ballistic jet with a parallel magnetic field configuration, model B2, is inconsistent with a flat spectrum. In this Section we therefore concentrate on model B1, a ballistic jet with a magnetic field perpendicular to the jet axis. With the assumptions made above equation (\ref{reduced}) reduces to 
\begin{equation}
F_{\nu} \sim 5.9 \times 10^{-7} \delta _{\mp}^2 x_0 r_0 D^{-2}B_0^{-1/2}\nu^{5/2} {l'_{\rm max}}^{5/2} \, {\rm mJy}
\end{equation}
for the ballistic jet model B1. Here and in the following, all quantities are measured in SI units unless indicated otherwise. Substituting the measurements discussed above we get an expression for $r_0$ as a function of $B_0$. A second equation relating the same quantities can be found from substituting $\gamma _{\rm thick}$ from equation (\ref{gamthick}) for $\gamma_{\rm max}$ in equation (\ref{ldmax}) resulting in
\begin{equation}
l'_{\rm max} = 2.0 \times 10^{11} \nu^{-1} B_0^{9/13} \kappa _0^{4/13} r_0^{4/13} = 1,
\end{equation}
where we again made use of the observed quantities. For initial equipartition we can eliminate $\kappa _0$ and solve for $B_0$. We can then calculate all other model parameters and they are summarised in Table \ref{modpara}. The model spectrum is plotted as the dashed line in Figure \ref{illustration2}.

The maximum brightness temperature of the jet emission does not depend on $l$ for model B1 and from equation (\ref{bright}) we find $T_{\rm b, max} =1.1\times10^{10}$\,K. We can also calculate at what distance $l$ along the jet the jet material is dense enough so that relativistic induced Compton scattering becomes important. From equation (\ref{crit}) we find that this loss process would only play a role for $l < 1.5\times10^{-7}$ which is well inside $l_{\rm min}$. We therefore conclude that our model can be applied to the jet of Cygnus X-1.

The jet radius $r_0$ is very small. The ballistic jet model is conical and therefore we can define a jet opening angle as $\theta = 2 r_0 / x_0 = 5"$ which is much smaller than the observational upper limit of $2^{\circ}$ \citep{ssf01}. It is of course possible to assume a larger radius for the jet by dropping the assumption of equipartition. However, setting the jet radius equal to the observational upper limit would imply that the jet material is out of equipartition by several orders of magnitude. 

The physical limits of the jet flow, $l_{\rm min}$ and $l_{\rm max}$, can be derived from equation (\ref{depth}) by setting $\tau \left( l_{\rm min} \right) = \tau \left( l_{\rm max} \right) = 1$ for those frequencies at which the flat spectrum is required to break to the optically thick or thin regime, respectively.  In our example of Cygnus X-1 we used a lower break of just under 100\,MHz and an upper break of $3\times 10^{14}$\,Hz corresponding to a wavelength of 1\,$\mu$m. The resulting limit $l_{\rm min}=4.3\times10^{-5}$, associated with the break to optical thin conditions, corresponds to a distance of about 6000 Schwarzschild radii from the central black hole with a mass of 10\,M$_{\odot}$ \citep{gzp99}. Since for the ballistic jet model B1 $l_{\rm min}$ decreases linearly with the break frequency, the lower physical limit decreases to about 500 Schwarzschild radii if the flat spectrum extends to the near-UV. Clearly the determination of the high frequency break of the flat part of the spectrum can put interesting constraints on the distance from the central black holes at which jets become ballistic. Similarly the low frequency break constrains the overall extent of the ballistic jet flow. 

The observable extent of the jet flow depends linearly on the observing frequency. Since $l'_{\rm max} \left(8.4\,{\rm GHz} \right) = 1$, we would expect that $l'_{\rm max} \left( 15\,{\rm GHz} \right) \sim 0.6$. This prediction significantly exceeds the tentative extension of the Cygnus X-1 jet of $l'_{\rm max} \left( 15\,{\rm GHz} \right) \le 0.13$ reported in \citet{ssg98}. However, the observations at 15\,GHz were not simultaneous with those at 8.4\,GHz. 

The strength of the magnetic field, $B_0$, at $x_0$ implies a field strength of around 50\,T at the distance $l_{\rm min}$. If $l_{\rm min}$ is reduced because the flat spectrum extends to the near-UV, then the magnetic field in the jet has a strength of 500\,T about 500 Schwarzschild radii away from the black hole. Again this demonstrates that our model can provide useful constraints on the conditions at the very base of the observed jets despite them not being spatially resolved.

The strength of the magnetic field, $B_0$, and the constant in the expression for the density of relativistic electrons, $\kappa _0$, can be used to estimate the power of the jet. We find that the energy transport rate associated with the magnetic field and the relativistic particles alone is $5.2\times10^{25}$\,W. A further $1.1 \times 10^{25}$\,W is added by the kinetic energy of the electrons. Not surprisingly these numbers are comparable to the estimates using the \citet{bk79} model \citep{fpdtb99}. If there is a cold proton for every relativistic electron in the jet, then its energy transport rate in terms of kinetic energy is $1.1\times 10^{28}$\,W. This power is about one order of magnitude below the time-averaged energy transport rate of the Cygnus X-1 jet, recently estimated from the observed interaction of the jet with the surrounding ISM \citep{gfk05}. The flux density of the flat jet spectrum does not vary significantly between available observations. Unless the jet power varies considerably over timescales longer than the timespan since the first available radio observations of Cygnus X-1, then our estimate strongly suggests the presence of a proton-electron plasma in the jet. 

\subsubsection{Adiabatic jet models}

\begin{figure}
\centerline{
\includegraphics[width=8.45cm]{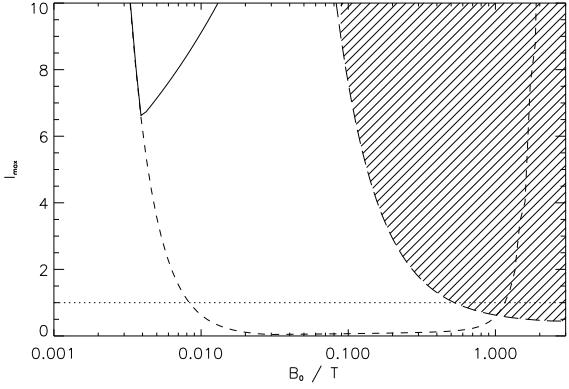}}
\caption{The maximum extent of the jet for an observing frequency of 8.4\,GHz in the adiabatic jet model A3 as a function of the magnetic field strength $B_0$. The solid line shows $l'_{\rm max}$ for initial equipartition. The dotted line indicates $l'_{\rm max} =1$ as required by the observations. The hatched region bounded by the long-dashed curve contains the allowed combinations of $B_0$ and $l'_{\rm max}$ for the assumption that $l_{\rm min}$ is larger than three Schwarzschild radii of a 10\,M$_{\odot}$ black hole and that the flat spectrum extends to the near-IR. The short-dashed line gives $l'_{\rm max}$ for an adiabatic jet out of equipartition with $f=10^{-6}$. The model parameters given in Table \ref{failedtab} are derived for the point where the short-dashed and dotted lines cross inside the hatched region.}
\label{lmaxest}
\end{figure}

All three adiabatic jet models are consistent with a partially flat emission spectrum. In fact, if we choose the geometrical parameter $a_1$ appropriate for a flat spectrum for each model, then the differences between the adiabatic models become small as the differences in the behaviour of the magnetic field as a function of $l$ between them is compensated for by the different degrees of confinement of the jet. Therefore and to simplify the discussion below, we focus on the adiabatic model A3 with an isotropic magnetic field. None of the conclusions change greatly for models A1 and A2. 

For the adiabatic jet model A3 our assumptions with equation (\ref{reduced}) lead to
\begin{equation}
F_{\nu} \sim 8.8 \times 10^{-7} \delta _{\mp}^2 x_0 r_0 D^{-2}B_0^{-1/2}\nu^{5/2} {l'_{\rm max}}^{5/3} \, {\rm mJy},
\end{equation}
where we have set $a=1/5$ to allow for a flat section in the spectrum. The calculation of $\gamma _{\rm max}$ now requires $l_{\rm thick}$, which must be determined from the implicit equation (\ref{gamadiabat}) for a given magnetic field strength, $B_0$. Figure \ref{lmaxest} demonstrates that the allowed minimum for $l'_{\rm max}$ exceeds unity for the assumption of initial equipartition. Since observations require $l'_{\rm max} =1$ at $8.4$\,GHz, the adiabatic models are incompatible with observations unless we drop the requirement of equipartition.

We now introduce a reduction factor $f$ such that the initial energy density of the relativistic electrons is a fraction $f$ of the initial energy density of the magnetic field. An example of the results for $l'_{\rm max}$ for $f=10^{-6}$ is shown in Figure \ref{lmaxest}. There are now two possible solutions for the strength of the magnetic field. However, we also require that the lower size limit of the jet, $l_{\rm min}$, accommodates a break of the flat spectrum to the optically thin regime in the near-IR. This lower limit cannot lie inside the last stable orbit of the central black hole. Thus we obtain another constraint on the solution because $l_{\rm min} \ge 3 R_{\rm S}$, where $R_{\rm S} = 3 \times 10^4$\,m is the Schwarzschild radius of a 10\,M$_{\odot}$ black hole. The additional constraint is also plotted in Figure \ref{lmaxest}. Only one of the two possible solutions for $f=10^{-6}$ is consistent with this constraint. It is also interesting to note that the solution requiring initial equipartition is also inconsistent with a physical meaningful value of $l_{\rm min}$. 

\begin{figure}
\centerline{
\includegraphics[width=8.45cm]{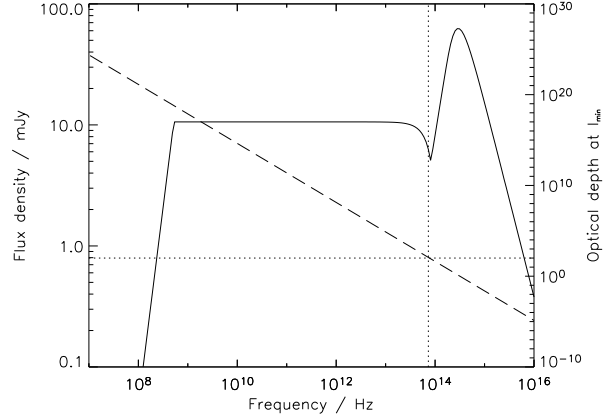}}
\caption{Spectrum of the adiabatic jet model A3 without initial equipartition using the parameters in Table \ref{failedtab}. The solid line shows the spectrum. The dashed line is the optical depths, $\tau_{\rm max}$, at $l_{\rm min}$. The dotted lines indicate the frequency for which $\tau_{\rm max} = 100$. As expected, the peak in the spectrum occurs at $\tau_{\rm max} \sim 1$.}
\label{failedspec}
\end{figure}

Using the remaining solution for $f=10^{-6}$ as an example, we can compute all remaining model parameters which are summarised in Table \ref{failedtab}. Figure \ref{failedspec} shows the resulting spectrum. It is flat over a wide range of frequencies with the required flux level. The dip at around $10^{14}$\,Hz is a result of the diminishing optical depths of the jet to radiation emitted by electrons with the limiting Lorentz factor $\gamma _{\rm max} = \gamma_{\rm thick} \left( l / l_{\rm thick} \right)^{-2 a_1/3}$. For higher frequencies those parts of the jet close to $l_{\rm min}$ which still contain electrons with Lorentz factors in excess of $\gamma _{\rm thick}$ contribute to the emission and cause the peak. The position of the peak is located at the frequency with optical depth $\tau_{\rm max} \sim 1$ at $l_{\rm min}$. Clearly the spectrum is not flat from radio to near-IR frequencies because of the emission peak. A flat spectrum extending over the entire radio to near-IR range could be achieved by moving the peak to higher frequencies. However, an appropriate adjustment of the model parameters would also tighten the constraints on the reduction factor $f$ and thereby on $B_0$.

\begin{table}
\begin{tabular}{lc}
Model parameter & Value\\
\hline
$x_0$ & $47$\,AU\\
$r_0$ & $1.3 \times 10^9$\,m\\[1ex]
$p$ & $2.5$\\
$B_0$ & $1.2$\,T\\[1ex]
$\kappa _0$ & $7.6\times10^{-8}$\,J$^{1.5}$\,m$^{-3}$\\
$D$ & $2$\,kpc\\[1ex]
$l_{\rm min}$ & $2.0\times 10^{-8}$\\
$l_{\rm max}$ & $200$\\[1ex]
$\gamma_{\rm max} \left( t_{\rm min} \right)$ & $10^6$\\
$v_{\rm j}$ & $0.97\,c$\\
$\vartheta$ & $40^{\circ}$\\
\hline
\end{tabular}
\caption{Parameters for the adiabatic jet model A3 without initial equipartition. The model spectrum is shown in Figure \ref{failedspec}.\label{failedtab}}
\end{table}

The maximum brightness temperature of the adiabatic model used here arises at $l_{\rm min}$ and is with $7.3\times 10^9$\,K well below the limit for efficient Compton scattering. The distance at which relativistic induced Compton scattering would become important in the adiabatic jet is $l=1.7\times 10^{-16}$. Again we are justified to neglect energy losses due to Compton scattering.

The jet radius, $r_0$, for the adiabatic jet is larger than for the ballistic jet. Formally, we cannot define a jet opening angle for adiabatic jets discussed here, because their shape is not conical. However, if a conical shape was assumed, then the opening angle inferred from the radius at $x_0$ would be 1.3'.

In our example, the lower limit of the physical extent of the jet, $l_{\rm min}$, corresponds to 4.7\,$R_{\rm S}$. While this is close to the theoretical limit and gravitational redshift would affect the spectrum, a further decreased value of the reduction factor would increase this limit at the expense of requiring an even stronger magnetic field. For adiabatic jets the constraints on the nature of the jet flow extend to even closer distances from the central black hole than in the ballistic case. 

For the adiabatic jet model A3 the observable extent of the jet flow is proportional to $\nu^{-3 / \left( 8 a \right)}$. In our example this relation implies that $l'_{\rm max} \left( 15\,{\rm GHz} \right) \sim 0.3$, which is still larger than the observed value of \citet{ssg98}, but closer than the prediction of the ballistic model. However, as mentioned above, the observations are not simultaneous and that may explain the discrepancy in both cases.

The strength of the magnetic field of 1.2\,T is high at $x_0$. However, because of the much more collimated geometry of the jet, the field strength only increases to 140\,T at $l_{\rm min}$. Nevertheless, the energy transport rate of the jet due to the magnetic field is very large with $10^{34}$\,W determined at $l_{\rm min}$. Due to the small value for the reduction factor $f$, the contribution of the relativistic electrons to any energetic considerations is negligible, even if the kinetic energy of possibly associated protons is taken into account. The derived jet power exceeds by far all previous estimates and is inconsistent with the time-averaged jet power \citep{gfk05}, unless the jet flow is suppressed for long periods on very long timescales. Note also, that this jet power corresponds to one hundred times the Eddington limiting luminosity of a 10\,M$_{\odot}$ black hole. The large jet power is caused by the significant reduction in the radiative efficiency of the synchrotron process well away from equipartition conditions.

\section{Conclusions}
\label{conc}

We construct a model for the synchrotron emission of partially self-absorbed jets. The model does not invoke a re-acceleration process for the relativistic electrons. All electrons are accelerated only once at the lower physical limit of the jet, $l_{\rm min}$. It is not necessary to postulate an unknown re-acceleration mechanism as was suggested in previous work \citep{bk79}, because synchrotron self-absorption counteracts excessive energy losses. 

Two classes of models are considered. The ballistic jets have a conical geometry and their contents do not suffer adiabatic energy losses. This situation may arise when jets are initially highly overpressured with respect to their environments. They expand unimpeded and random thermal energy is converted into bulk kinetic energy, but not dissipated to any external medium. At the end of this very rapid expansion the mean free path of the jet material exceeds the physical dimensions of the jet itself and follows ballistic trajectories. 

The other class of models considered here are adiabatic jets confined by either the pressure of the gas surrounding them, surface magnetic fields or both. Their geometrical shape is dictated by the details of the confinement mechanism which is not the subject of this paper. The jet material dissipates energy to the external gas during its adiabatic expansion. 

Both classes of models can predict flat emission spectra if energy losses of individual electrons are neglected. The ballistic jet with a magnetic field perpendicular to the jet axis produces a flat spectrum without further assumptions. The adiabatic jet models require a specific jet geometry to allow for flat emission spectra. 

Both model classes can predict flat emission spectra, even when taking energy losses of the electrons and the magnetic field into account. Synchrotron self-absorption prevents radiative energy losses below a critical Lorentz factor $\gamma _{\rm thick}$. In the ballistic case $\gamma _{\rm thick}$ is constant along the jet flow and because adiabatic energy losses are absent, the energy distribution of the relativistic electrons remains stationary. The emission properties of the jet in this case are essentially identical to the model of \citet{bk79}. For the adiabatic jet the with a perpendicular magnetic field electrons with Lorentz factors at and below $\gamma _{\rm thick}$ continue to lose energy because of the sideways expansion of the jet. However, for a geometrical shape given by $r \propto x^{1/5}$ the resulting spectrum is again flat. Other configurations of the magnetic field can also lead to flat emission spectra. The spectral slope is very sensitive to the jet shape. For example, slightly changing the jet shape to $r \propto x^{1/4}$ results in a spectrum with $F_{\nu} \propto \nu^{0.38}$. 

We show an application of the model to observations of a resolved jet in the X-ray binary Cygnus X-1 \citep{ssf01}. As input we use the flux density of the jet in the flat part of the spectrum, the physical extent of the resolved jet and an assumed break of the flat spectrum to optically thin conditions in the near-IR. Both model classes can be made consistent with the observational constraints. However, in doing so the adiabatic jet models require a significant departure from energy equipartition between the magnetic field and the relativistic particles. The associated reduced radiative efficiency of the jet plasma implies extremely high energy transport rates for the jet of around $10^{34}$\,W. This jet power exceeds the Eddington limiting luminosity of a 10\,M$_{\odot}$ black hole by two orders of magnitude. 

The ballistic jet is consistent with current observations and requires energy transport rates well below the time-averaged jet power \citep{gfk05}. This result holds even if the kinetic energy of one non-radiating proton per relativistic electron is taken into account. Unless Cygnus X-1 ceases to produce a jet for long periods of time, then the ballistic model requires a proton-electron jet plasma to explain the large accumulated energy in the region where the jet interacts with the surrounding gas. 

The jet opening angle of 5" required by the ballistic model is very small. The velocity perpendicular to the jet axis established in the jet acceleration region must be very small for this model to work. If the ballistic jet results from a rapid initial expansion of a highly overpressured jet, then the requirements are even more severe. This result places stringent limits on the acceleration process. 

Both model classes can probe the conditions in jets on scales unresolvable with current telescopes. The most important measurement in this respect is the high frequency cut-off of the flat spectrum which is formed closest to the jet acceleration region. In order to resolve the remaining uncertainties with the model, the most helpful measurements would be to resolve the jet along its axis in quasi-simultaneous observations at two different frequencies. As mentioned above, even a factor 2 between the observing frequencies employed would help to constrain the geometrical shape of the jet and thereby help to decide which of the model classes is compatible with observations. 

Obviously from the work presented here we cannot rule out the possibility of continued re-acceleration of the radiating electrons in the jet. However, such an energetisation process is not necessary to produce flat spectra from self-absorbed, synchrotron emitting jets.

\section*{Acknowledgments}

It is a pleasure to thank E. Gallo, J. Miller-Jones and R.P. Fender for helpful discussions. I also thank the referee A. Marscher for many helpful comments that improved the paper.

\def\newblock{\hskip .11em plus .33em minus .07em}

\bibliography{crk}
\bibliographystyle{mn2e}

\end{document}